# $\mu$PIC（Micro-Pixel Chamber）を用いた中性子イメージング検出器の開発

パーカー ジョセフ ドン [1]，原田 正英 [2]，服部 香里 [1]，岩城 智 [1]，株木 重人 [1]，
岸本 祐二 [1]，窪 秀利 [1,4]，黒澤 俊介 [1]，松岡 佳大 [1]，身内 賢太朗 [1]，水本 哲矢 [1]，
西村 広展 [1]，奥 隆之 [2]，澤野 達哉 [1]，篠原 武尚 [2]，鈴木 淳市 [2]，高田 淳史 [1]，
谷森 達 [1]，上野 一樹 [1]，池野 正弘 [3,4]，田中 真伸 [3,4]，内田 智久 [3,4]
（京都大学大学院理学研究科 [1]，日本原子力研究開発機構 J-PARC センター [2]，
高エネルギー加速器研究機 素粒子原子核研究所 [3]，オープンソースコンソーシアム Open-It [4]）

# Development of a Time-resolved Neutron Imaging Detector Based on the $\mu$PIC Micro-Pixel Chamber

Joseph Don PARKER [1], Masahide HARADA [2], Kaori HATTORI [1], Satoru IWAKI [1],
Shigeto KABUKI [1], Yuji KISHIMOTO [1], Hidetoshi KUBO [1,4], Shunsuke KUROSAWA [1],
Yoshihiro MATSUOKA [1], Kentaro MIUCHI [1], Tetsuya MIZUMOTO [1],
Hironobu NISHIMURA [1], Takayuki OKU [2], Tatsuya SAWANO [1], Takenao SHINOHARA [2],
Jun-ichi SUZUKI [2], Atsushi TAKADA [1], Toru TANIMORI [1], Kazuki UENO [1],
Masahiro IKENO [3,4], Manobu TANAKA [3,4] and Tomohisa UCHIDA [3,4]
Department of Physics, Kyoto University [1],
J-PARC Center, Japan Atomic Energy Agency [2],
Institute of Particle and Nuclear Studies, KEK [3] and
Open Source Consortium (Open-It) [4]

ABSTRACT
We have developed a prototype time-resolved neutron imaging detector employing a micro-pattern gaseous detector known as the micro-pixel chamber ($\mu$PIC) coupled with a field-programmable-gate-array-based data acquisition system. Our detector system combines 100$\mu$m-level spatial and sub-$\mu$s time resolutions with a low gamma sensitivity of less than $10^{-12}$ and high data rates, making it well suited for applications in neutron radiography at high-intensity, pulsed neutron sources. In the present paper, we introduce the detector system and present several test measurements performed at NOBORU (BL10), J-PARC to demonstrate the capabilities of our prototype. We also discuss future improvements to the spatial resolution and rate performance.
Keywords: *Neutron imaging, Micro-pattern gaseous detector, Neutron resonance absorption spectroscopy, Neutron Bragg-edge transmission*

## 1. Introduction

The micro-pixel chamber ($\mu$PIC), a micro-pattern gaseous detector, has been developed in our group at Kyoto University [1]. The $\mu$PIC, manufactured by DaiNippon Printing Co., Ltd., using standard printed circuit board techniques, features a 400-$\mu$m pitch and two-dimensional strip readout (Fig. 1), coupled with a fast, compact field programmable gate array (FPGA) based data acquisition system. Detectors incorporating the $\mu$PIC have already been successfully developed for applications ranging from MeV gamma-ray astronomy [2] to medical imaging [3] to small-angle X-ray scattering [4]. By adding a small amount $^3$He to the gas mixture, the $\mu$PIC becomes an effective thermal neutron area detector [5].

Using the $\mu$PIC, we have developed a prototype time-resolved neutron imaging detector with 100$\mu$m-level spatial and sub-$\mu$s time resolutions, excellent gamma rejection, and high-rate capability [6,7]. The detector is well suited for use at pulsed neutron sources where imaging is combined with neutron energy measured via time-of-flight (TOF). The prototype employs a time-projection-chamber, consisting simply of a drift cage and 10 × 10-cm$^2$ $\mu$PIC, contained within an aluminum vessel (Fig. 2). Neutrons are detected via absorption on $^3$He with 18% efficiency at 25.3 meV using Ar-$C_2H_6$-$^3$He (63:7:30) gas at 2-atm total pressure and a gas depth of 2.5 cm. The fast, FPGA-based data acquisition system records both the energy deposition (via time-over-threshold, or TOT, without the added expense and complexity of ADCs) and 3D tracking information (2D position plus time) for each neutron event.

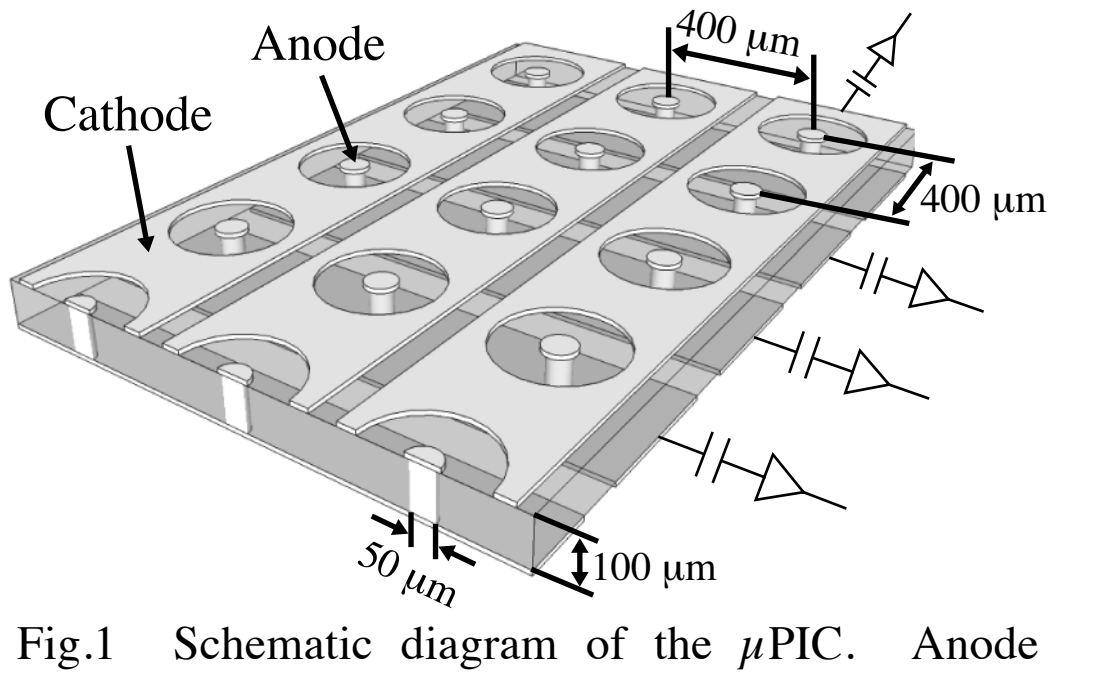

Fig.1 Schematic diagram of the $\mu$PIC. Anode pixels connected on the backside by copper strips with orthogonal cathode strips on the topside produce a two-dimensional readout.

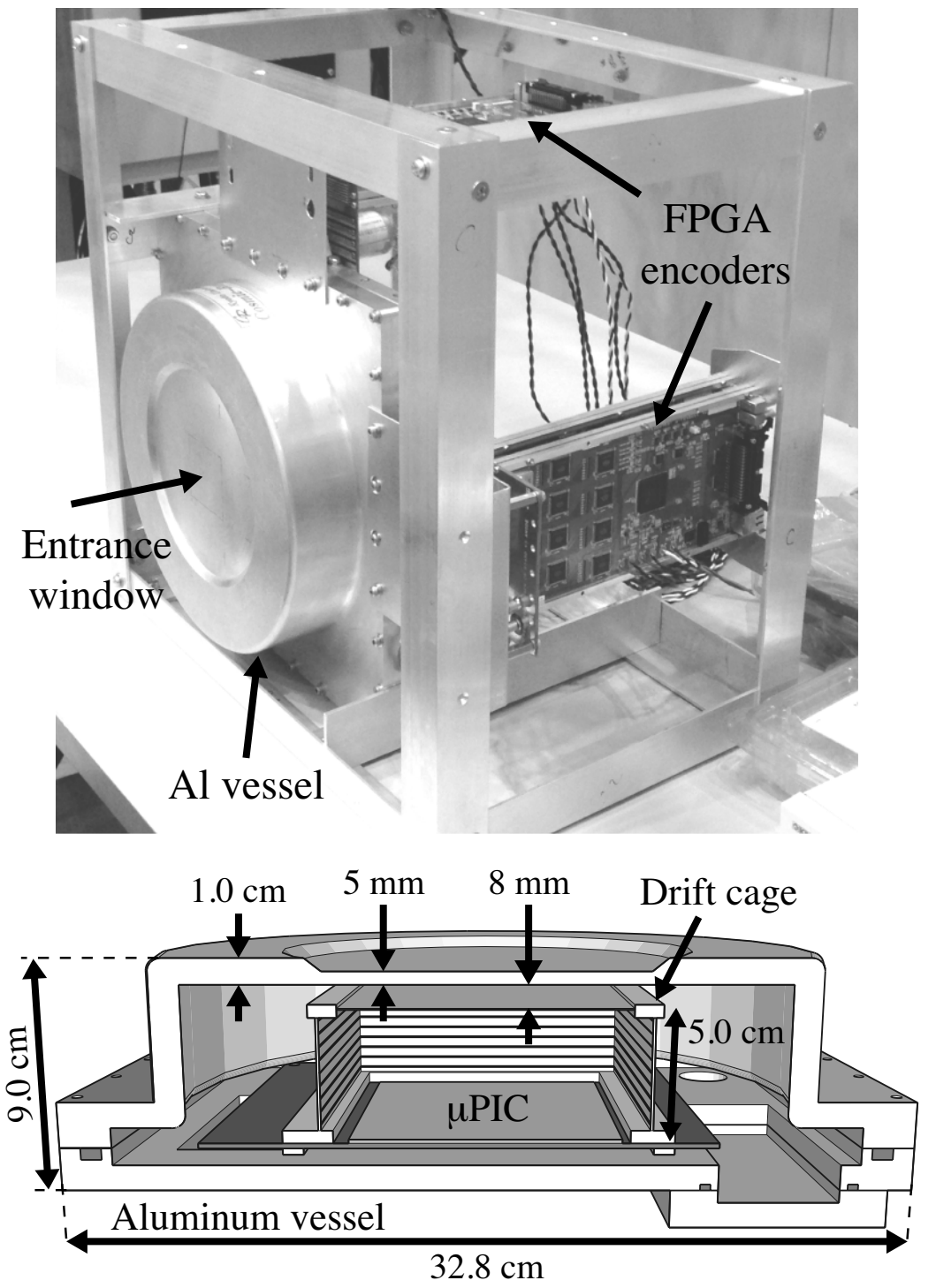

Fig.2 Photograph of the neutron imaging detector prototype and diagram showing a cut-away of the detector vessel with a 5-cm drift cage (a 2.5-cm drift cage is currently installed).

Our detector currently achieves a spatial resolution of 100 to 120 $\mu$m, time resolution less than 0.6 $\mu$s, and gamma sensitivity below $10^{-12}$. Preliminary testing of a recent upgrade to the data acquisition hardware showed an increase in neutron count rate from 200 to 700 kcps (kilo-counts-per-second), with ~3 Mcps expected after optimization of the FPGA code and gas mixture. Additionally, the detector can operate for ~2 years per gas filling, lowering the consumption of $^3$He gas (one filling requires only 1 to 2 liters). The good time resolution and high-rate capability make our detector particularly suited to TOF-based techniques such as neutron resonance absorption spectroscopy [8] and Bragg-edge transmission [9,10].

In the present paper, we introduce our detector and show the results of several test measurements, followed by a discussion of recent and future improvements to the detector. All data analysis was carried out using custom software based on the ROOT object-oriented framework [11].

## 2. Operating principle

The absorption of low energy neutrons on $^3$He produces a proton-triton pair with total kinetic energy of ~764 keV and a combined track length of ~8 mm in the 2-atm gas of our detector. The range of the proton is about three times that of the heavier triton, making the separation of the two particles essential for an accurate determination of the neutron interaction position. As the proton and triton emerge essentially back-to-back, tracking alone is not sufficient, and additional information, in the form of the energy deposition (estimated via time-over-threshold), must be utilized to perform the separation.

This is illustrated in Fig. 3, which shows a neutron interaction event measured with our prototype system. Fig. 3(a) shows the track, given by the relative time and strip number of each hit, and Fig. 3(b) shows the associated TOT distribution. While tracking alone gives no indication of the neutron interaction point, the TOT distribution clearly displays the Bragg peaks of the proton and triton, facilitating their separation.

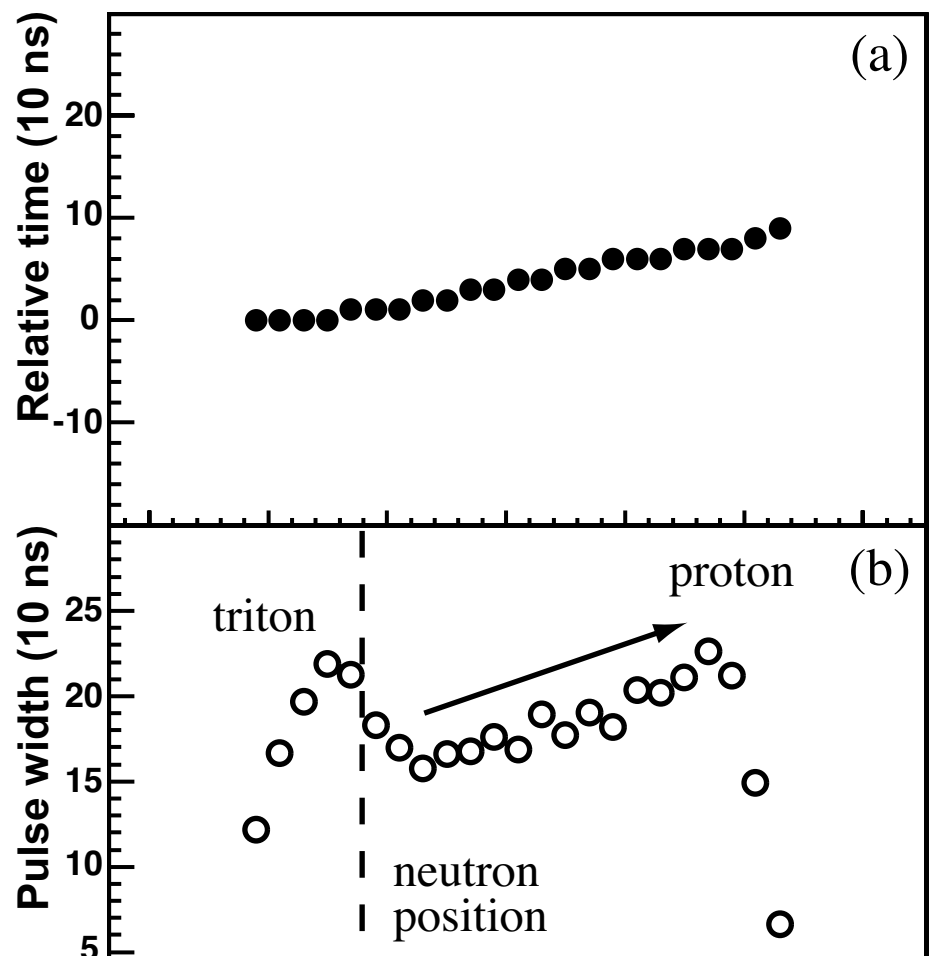

Fig.3 Measured proton-triton track. (a) Position (strip number) versus time and (b) TOT for each hit strip are shown. The neutron position (dashed line) and the slow rise of the proton Bragg peak (arrow) are indicated in (b).

In addition to aiding in the neutron position reconstruction, the combination of the 3D tracking and energy deposition via TOT provides a powerful means for rejecting background on an event-by-event basis, achieving an effective gamma sensitivity of less than $10^{-12}$ (95% confidence level) while maintaining stable, robust neutron identification. Such a low sensitivity is possible due to the fact that neutron interactions in our detector produce proton-triton tracks with similar track lengths and energy deposition, while gammas and protons scattered by fast neutrons produce a range of both. Additionally, the characteristic shape of the TOT distributions for proton-triton tracks allows us to positively identify these events even when background events overlap in track length and/or total energy deposition. Further discussion of the background rejection and the determination of the gamma sensitivity are given in Ref. [6].

## 3. Demonstration measurements

We have carried out several test experiments at NOBORU (NeutrOn Beamline for Observation and Research Use), BL10 at the Materials and Life Science Experimental Facility (MLF) located within the Japan Proton Accelerator Research Complex (J-PARC) [12]. Examples of radiography, neutron resonance absorption, and Bragg-edge transmission measurements are given below.

### 3.1 Neutron radiography

For imaging, the neutron position is found by fitting the measured TOT distributions event-by-event with the expected distributions determined using a GEANT4 [7,13] simulation of our detector system. Using this method, we achieved a spatial resolution of 100 to 120 $\mu$m, with the lower value obtained by applying strict tracking quality cuts at a cost of about 68% of neutron events. A detailed study of the spatial resolution can be found in Ref. [7].

Fig. 4 shows images of a wristwatch (February 2011) and the sensitivity indicator (SI) and beam purity indicator (BPI) described in Ref. [14] (March 2012). The images, with ~120 $\mu$m resolution, were obtained at NOBORU using a collimator ratio (L/D) of 1875 and measurement times of 29 (wristwatch) and 25 (indicators) minutes. In particular, the seven vertical slits of the SI, with widths from 250 down to

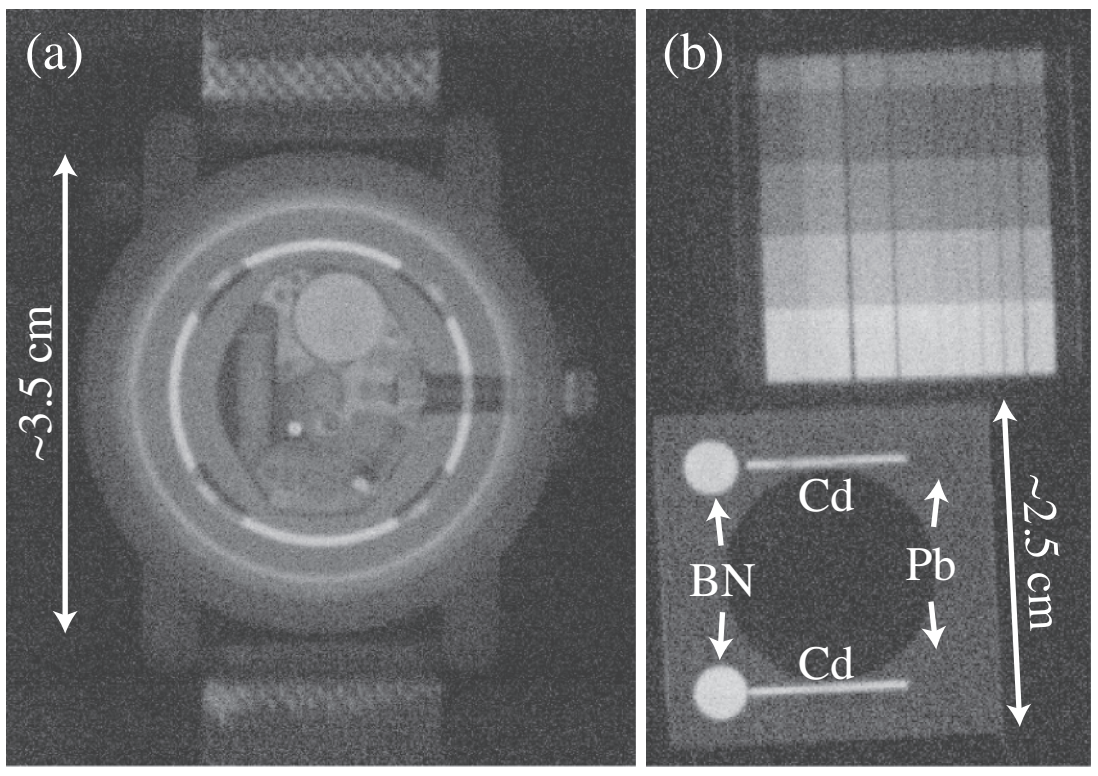

Fig.4 Radiographic images taken at NOBORU. Images of a) a wristwatch, and b) sensitivity (upper) and beam purity (lower) indicators.

13 $\mu$m, show the spatial resolution, while the virtual invisibility of the Pb disks of the BPI is a consequence of the excellent gamma rejection of our detector.

### 3.2 Neutron resonance absorption

Neutron resonance absorption techniques take advantage of the tendency of some nuclides to preferentially absorb neutrons at specific energies, unique to each nuclide, to study isotopic composition and temperature [8]. Resonance absorption can be directly observed by measuring the neutron transmission of the sample, $Tr(TOF)$, calculated as:

$$Tr(TOF) = \frac{I(TOF)}{I_0(TOF)} \tag{1}$$

where $I(TOF)$ and $I_0(TOF)$ are the neutron intensities measured with and without the sample, respectively. Due to the preferential absorption, the transmission will show sharp drops near a resonance, typically occurring for neutrons with energies above 1 eV, or a TOF of less than 1 ms at NOBORU.

Fig. 5 shows the preliminary results of a computed tomography (CT) measurement performed at NOBORU in December 2012. The sample, consisting of Cu, Mo, W, Ag, and In rods with diameters from 2 to 6 mm (Fig. 5(a)), was measured at 16 angles (~14 × $10^6$ neutrons per angle) with the axis of rotation aligned along the length of the rods, and images of a single CT slice were reconstructed (Fig. 5 (b—f)). The indicated neutron TOF cuts, chosen to select resonances at 2039 and 579 eV (Cu), 45 eV (Mo), 19 eV (W), 5.2 eV (Ag), and 1.5 eV (In), clearly enhance the individual elements in each image.

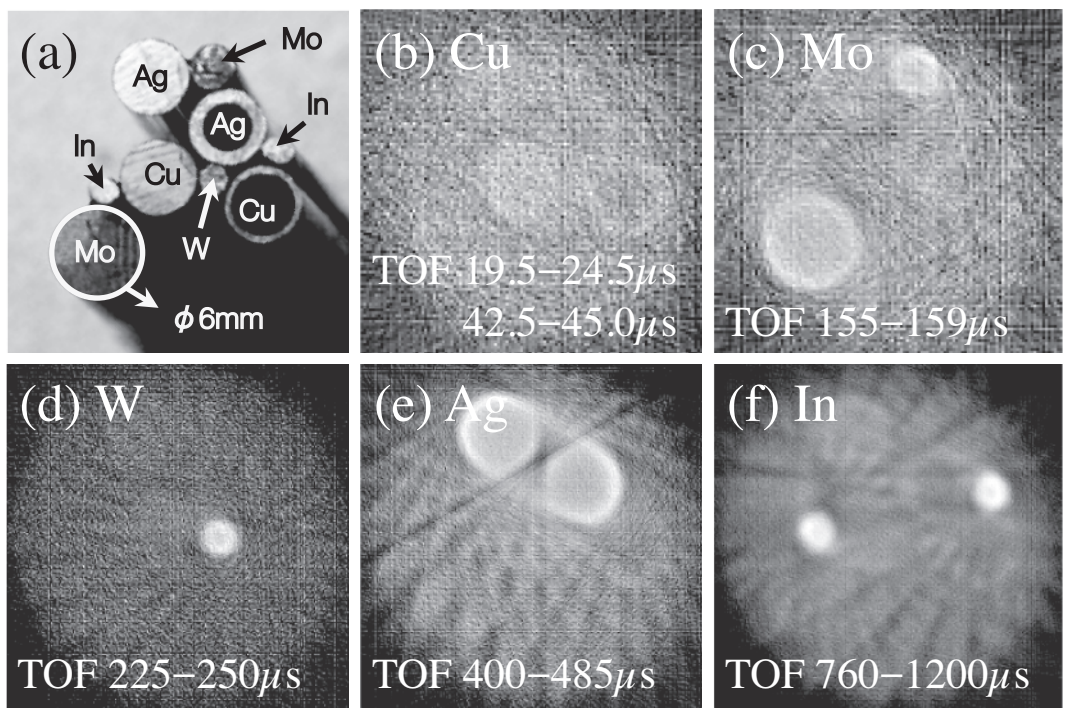

Fig.5 Resonance CT. a) Photo of sample from above, and b—f) computed tomography images of specific elements selected using neutron TOF (b: Cu, c: Mo, d: W, e: Ag, and f: In).

### 3.3 Bragg-edge transmission

Bragg edges appear as jumps in the transmission at the point where the neutron wavelength exceeds the Bragg-scattering condition for a particular set of crystal planes, and are observed for energies down to a few meV (or for a TOF up to about 20 ms at NOBORU). From the position (in wavelength) of the Bragg edge, one can determine the crystal spacing ($d = \lambda/2$), and by studying the shape of the spectrum in the vicinity of the edge, one can learn about crystal properties such as grain size and texture [9,10].

As one example, Bragg-edge transmission and a TOF-capable area detector such as ours can be used to make a 2D map, in a single measurement, of the component of strain along the beam direction. This is achieved by using the crystal as a microscopic strain gauge, with the strain component, $\varepsilon$, given by:

$$\varepsilon = \frac{d - d_0}{d_0} \quad (2)$$

where $d$ ($d_0$) is the strained (unstrained) crystal spacing. By taking six such measurements, the entire strain tensor can be reconstructed.

Fig. 6 shows preliminary results for a welded steel sample measured at NOBORU in June 2010. The sample, consisting of a 1-cm thick section of two 316L stainless steel plates joined by a double TIG weld, is shown in Fig. 6(a), along with a schematic indicating the beam (and strain) axis. Fig. 6(b) shows the Bragg-edge spectra for the weld area and each steel plate, with edge spacing consistent with an FCC (Face-Centered-Cubic) crystal structure. Fig. 6(c) shows the estimated strain distributions for the upper and lower weld regions. The measured strain indicates the weld is under tension (i.e., it has a higher value of strain relative to the steel plates). See Ref. [15] for details about the measurement and data analysis.

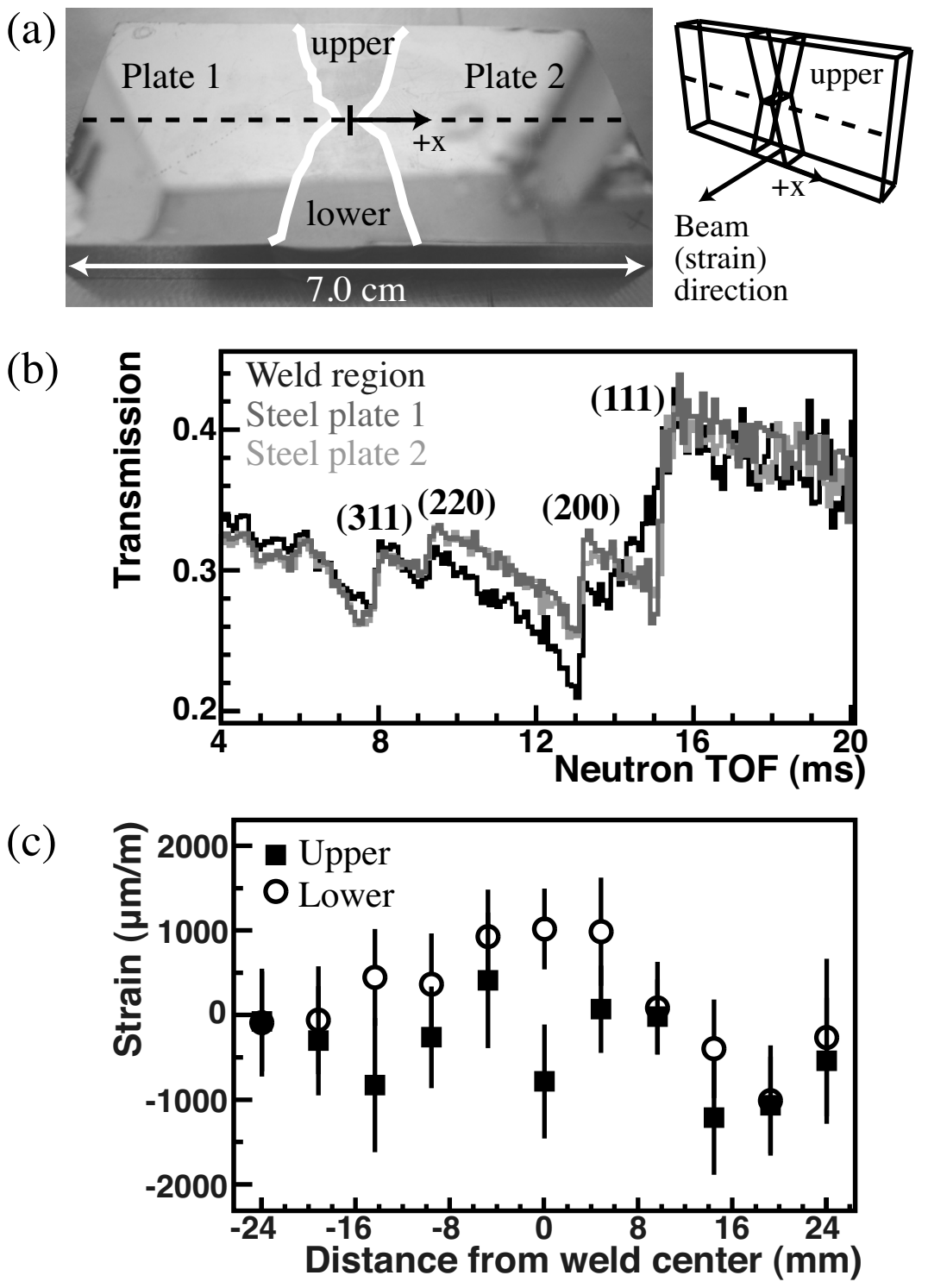

Fig.6 Strain scanning via Bragg-edge transmission. Shown here: a) welded steel sample (weld area indicated by white lines), b) neutron TOF spectra for weld region and steel plates (with Miller indices of prominent edges), and c) longitudinal strain for upper and lower weld regions.

## 4. Detector improvements

Recently, we have upgraded our data acquisition hardware, replacing a bulkier system originally developed some 10 years ago for use in low-rate, Compton gamma imaging $\mu$PIC applications [16,17] (Fig. 7, upper) with a compact, modular data encoder combining new low-power ASICs (Application Specific Integrated Circuits) and an FPGA on a single board (Fig. 7, lower). The new system was tested at NOBORU in February 2013, where we confirmed a maximum neutron rate of 700 kcps (compared to 200 kcps for the previous system). After further optimizing the FPGA logic, follow-up tests at Kyoto University using a $\mu$PIC X-ray imaging detector and X-ray generator indicated that a neutron rate of ~920 kcps can be achieved with the improved logic. Furthermore, by calculating TOT on the FPGA (currently being implemented) and increasing the stopping power of the gas, we can significantly reduce the amount of data generated by each neutron event and increase the counting rate to the order of 3 Mcps, as described in Ref. [7]. We also plan to upgrade the data transfer from the current VME (via 32-bit parallel transfer at 50 MHz) to SiTCP (via Gigabit Ethernet) to eliminate the loss of live time caused by the slow VME-to-PC transfer speed suffered by the current system and increase overall ease-of-use.

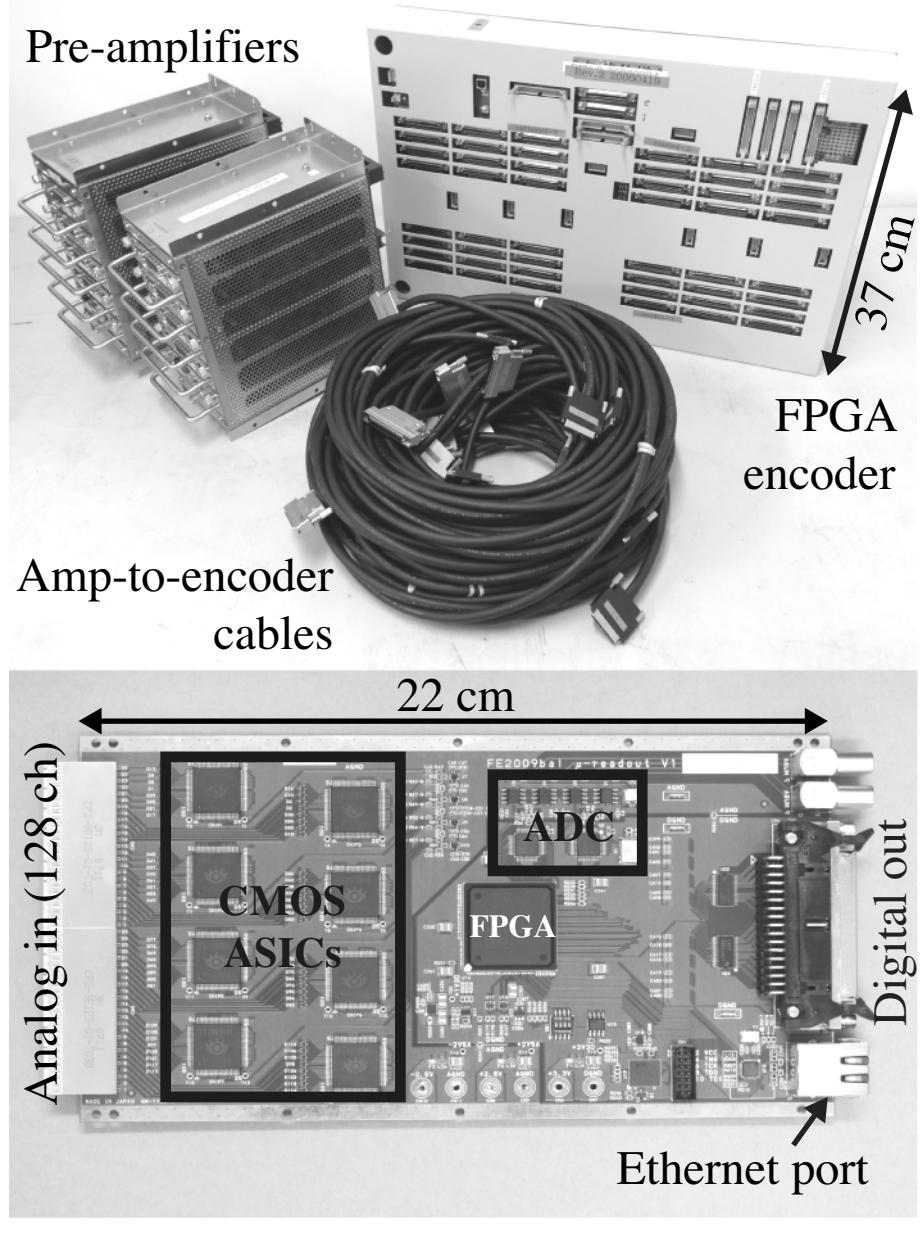

Fig.7 Photographs of previous data acquisition hardware (upper) and new compact, modular encoder (lower). Our prototype uses four new-type boards (for 256 anode and 256 cathode channels).

Using our detector simulation, we have also identified several ways to improve the spatial resolution, including: 1) optimization of the gas mixture for reduced electron diffusion and/or shorter track lengths, 2) reduction of the gain variation of the $\mu$PIC through tighter manufacturing controls, and 3) reduction of the strip pitch of the $\mu$PIC. We estimate a reduction of ~45% in the spatial resolution (i.e., to less than 60 $\mu$m) after making these improvements to the detector [7]. In addition, we can increase the efficiency of our detector by increasing the fraction of $^3$He included in the final gas mixture. Table 1 lists the operating parameters of our current prototype detector along with those expected after optimization.

Table 1 $\mu$PIC operating parameters for current prototype and those expected after optimization.

| Parameter | $\mu$PIC | $\mu$PIC (optimized) |
|---|---|---|
| Spatial resolution | 100–120 $\mu$m | ~60 $\mu$m |
| Time resolution | < 0.6 $\mu$s | |
| Efficiency @25.3meV | 18 % | > 30 % |
| Counting rate (counts/s) | > 700 kcps | ~3 Mcps |
| Gamma sensitivity | < $10^{-12}$ | |
| Typical area | 100 – 900 cm$^2$ | |

5. Conclusion

We have developed a time-resolved neutron imaging detector which achieves 100-$\mu$m spatial resolution, 0.6-$\mu$s time resolution, gamma sensitivity less than $10^{-12}$, and neutron counting rate > 700 kcps. Further improvements to the detector, as outlined here, should yield a spatial resolution of ~60 $\mu$m and a neutron counting rate of nearly 3 Mcps. Furthermore, the simple structure of our detector in conjunction with the flexibility of printed circuit board technology allows the possibility of adopting a modular structure, with each individual module operating at ~3 Mcps. We have also shown that by combining high-resolution imaging with event-by-event TOF measurement, our detector becomes a powerful and flexible tool for radiographic studies at pulsed neutron sources.

Acknowledgements

This work was supported by the Quantum Beam Technology Program of the Japan Ministry of Education, Culture, Sports, Science and Technology (MEXT). The neutron experiments were performed at NOBORU (BL10) of the J-PARC/MLF with the approval of the Japan Atomic Energy Agency (JAEA), Proposal No. 2009A0083. The authors would like to thank the staff at J-PARC and MLF for their support during our test experiments.

Joseph Don PARKER（じょせふ　どん　ぱーかー）
略歴：2005, Carnegie Mellon University, Department of Physics, Ph.D.
2006—2008，京都大学大学院理学研究科，物理第二教室，原子ハドロン研究室，外国人特別研究員
2008 より，京都大学大学院理学研究科，物理第二教室，宇宙線研究室，研究員

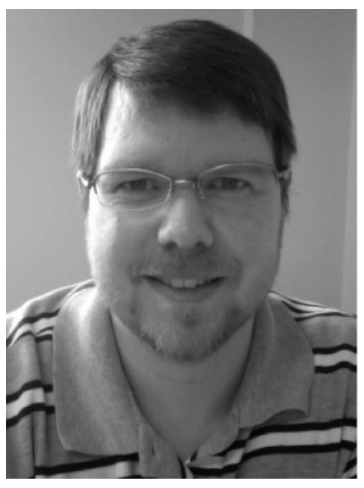

所属：京都大学大学院理学研究科
e-mail：jparker@cr.scphys.kyoto-u.ac.jp
専門：検出器開発
趣味：ギター，フォタグラフィ